# Crystal phase engineering of self-catalyzed GaAs nanowires using RHEED diagram

T. Dursap[1], M. Vettori[1], A. Danescu[1], C. Botella[1], P. Regreny[1], G. Patriarche[2], M. Gendry[1], J. Penuelas[1]

[1] Institut des Nanotechnologies de Lyon-INL, UMR 5270 CNRS, Université de Lyon, École Centrale de Lyon, 36 avenue Guy de Collongue, F-69134 Ecully cedex, France

[2] Centre de Nanosciences et de Nanotechnologies-C2N, UMR 9001 CNRS, Université Paris-Sud - Université Paris-Saclay, 10 Boulevard Thomas Gobert, 91120 Palaiseau, France

* To whom correspondence should be addressed. E-mail: jose.penuelas@ec-lyon.fr

**ABSTRACT:** It is well known that the crystalline structure of the III-V nanowires (NWs) is mainly controlled by the wetting contact angle of the catalyst droplet which can be tuned by the III and V flux. In this work we present a method to control the wurtzite (WZ) or zinc-blende (ZB) structure in self-catalyzed GaAs NWs grown by molecular beam epitaxy, using *in situ* reflection high energy electron diffraction (RHEED) diagram analysis. Since the diffraction patterns of the ZB and WZ structures differ according to the azimuth $[1\bar{1}0]$, it is possible to follow the evolution of the intensity of specific ZB and WZ diffraction spots during the NW growth as a function of the growth parameters such as the Ga flux. By analyzing the evolution of the WZ and ZB spot intensities during some NW growths with specific changes of Ga flux, it is then possible to control the crystal structure of the NWs. ZB GaAs NWs with a controlled

WZ segment have thus been realized. Using a semi-empirical model for the NW growth and our *in situ* RHEED measurements, the critical wetting angle of the catalyst droplet for the structural transition is deduced.

1. Introduction

Semiconductor nanowires (NWs) are greatly promising materials for future nanoelectronic and nanophotonic devices[1–5]. The fabrication of these NWs is mainly based on the vapor-liquid-solid (VLS) growth mechanism, a method where atoms are transported from the vapor phase to the NW solid phase through a liquid catalyst droplet[6]. The occurrence of two crystal phases in non-nitride III-V NWs has attracted considerable attention in the last decade since the model developed by F. Glas *et al.* in 2007 to explain the nucleation of the Wurtzite (WZ) or the Zinc Blende (ZB) crystal phase in such NWs[7]. Indeed, while only the ZB phase is reported in bulk III-V materials, III-V NWs exhibit the ZB or WZ phase depending on the growth parameters. It has been theoretically predicted that WZ GaAs has a slightly larger band gap than its ZB counterpart, with a positive conduction-band offset of up to 149 meV[8] and a slightly different band structure. Moreover, ZB and WZ phases are known to exhibit different electronic[9], optical[10–12], mechanical[13], thermoelectric[14] and piezoelectric properties[15,16]. This tunable physical properties only arising from the crystalline structure without incorporation of foreign chemical elements is particularly attractive for device fabrication. The possibility to control the crystal phase of GaAs NWs during the growth opens the possibility to develop a wide range of heterostuctured NWs including quantum dots, quantum disks as thin as a single monolayer (ML) and superlattices[17].

The WZ phase is often obtained for gold-catalyzed III-V NWs, while the ZB phase is mostly obtained for self-catalyzed ones. In the latter case, the WZ phase is however often observed at the NW top near the Ga droplet and can be ascribed to the end of the NW growth when the Ga flux is stopped and that the Ga droplet is consumed under the As flux[18–24]. The control of the ZB and WZ phases in self-catalyzed GaAs NWs has thus become a major challenge. Based on the models of F. Glas[7] and V. Dubrovskii[25], occurrence of the WZ or ZB phase in self-catalyzed GaAs NWs has been explained by the position of the nucleation for a new atomic layer either at the triple phase line (TPL) or inside the droplet, respectively[19,22,26–28]. It was recently shown that the nucleation position mainly depends on the droplet wetting angle and therefore on the catalyst droplet volume, for both Au-catalyzed and self-catalyzed GaAs NWs[29–31]. A critical wetting angle $\beta_c$ in the 121°-124° range for Au catalyzed NWs[29,31] and of about 127° for self-catalyzed NWs[30] has been experimentally observed for a transition from WZ to ZB crystal phase above this critical angle. From the *in situ* transmission electron microscopy (TEM) observations of Jacobsson *et al.*, above this angle a truncated facet is present at the NW top, thus determining the nucleation site inside the droplet and leading to the ZB crystal phase[29]. In self-catalyzed growth, the Ga droplet volume is mainly dependent on the growth conditions, in particular on the Ga and As fluxes. Many previous studies have thus reported on the control of the GaAs NW crystal phase by tuning the Ga and/or As fluxes[22,27,30,32–36]. In these studies the crystal phase characterizations have been mainly performed *ex situ* by TEM. However, an *in situ* and real-time characterization tool also appeared to be very useful in order to characterize and possibly tune the crystal phase of the self-catalyzed GaAs NWs during the growth. Compared to *in situ* X-ray diffraction[34] and *in situ* TEM[29,31], the reflection high energy electron diffraction (RHEED) technique is commonly coupled to molecular beam epitaxy (MBE) reactors to follow the structural properties of growing layers and nanostructures. Despite this, there are relatively few studies reporting on RHEED observations during the self-catalyzed GaAs NW growth. Scarpellini *et al.*[24] reported on the consumption of the Ga droplets at the end of the growth, while Rudolph *et al.*[21] reported on the influence of the As flux on the InAs NW structural properties and Bastiman *et al.* reported on the incubation time of GaAs NWs[37]. Only recently, *in situ* RHEED characterizations of the NW growth coupled with *ex situ* TEM measurements were reported by Jo *et al.*[38].

In this work, we focus on the characterization of the growth of self-catalyzed GaAs NWs on a Si(111) substrate using *in situ* RHEED. In particular, we aim to control the formation of the ZB or WZ crystal phase of the GaAs NWs as a function of the Ga flux amounts by using the RHEED pattern. TEM measurements of some NWs were performed to check the obtained crystal phases.

2. **Experiment**

All the samples were grown on epi-ready Si(111) substrates using a solid-source MBE reactor. The native oxide on the substrates was preserved to enable the self-catalyzed growth[39]. Each substrate was only cleaned in acetone and ethanol solutions for 10 min. The substrate was degassed at 200 °C in ultra-high vacuum and introduced inside the MBE reactor. One ML of Ga was pre-deposited at 480°C to form the Ga droplets[40,41]. The sample temperature was then increased to the NW growth temperature $T_G$=600°C and the growth was initiated by the simultaneous opening of the Ga and As fluxes. The MBE system was supervised by a homemade software which allows fine control of the Ga and $As_4$ flux, shutters and valves. For standard conditions, the NWs were grown with an $As_4$ flux of 1.15 ML/s and a Ga flux 0.5 ML/s corresponding a V/III flux ratio = 2.3. RHEED measurements were systematically performed with an electron beam energy set to 30 keV to obtain information on the crystal structure of the NWs during the growth process. The sample rotation was systematically stopped during the recording of RHEED patterns. The samples were then observed with a JEOL scanning electron microscope (SEM) using an acceleration voltage of 10 kV and TEM measurements were performed on a FEI Titan Themis 200 working at 200 kV.

## 3. Results

### 3.1. GaAs NW crystal structure

A typical RHEED pattern measured along the [1$\bar{1}$0] azimuth of self-catalyzed GaAs NWs when both WZ and ZB phases are present is shown in Figure 1. The position of the spots is in agreement with an epitaxial growth of NWs on Si(111): the GaAs [111] and [1$\bar{1}$0] axis are parallel to the Si [111] and [1$\bar{1}$0] axis, respectively. The corresponding spots of ZB and WZ phases were indexed. The (002)t ZB and (10-12) WZ spots whose intensities will be measured during growth are indicated by the green and red arrows, respectively. The SEM picture of Figure 2(a) shows the NW morphology after 10 min of growth: their length is about 800 nm and diameter in the 40-60 nm range with density close to 1 NW/µm². Some parasitic nanocrystals among the NWs are evidenced; however the substrate surface is still visible since the growth temperature was optimized in order to minimize this parasitic growth. Figure 2(b) shows the typical RHEED pattern obtained along the [1$\bar{1}$0] azimuth during the NW growth. Only ZB spots are observed indicating the growth of pure ZB NWs with twin planes. Figure 2(c) shows the RHEED pattern at the end of the NW growth after closing the Ga shutter and cooling the sample under the As$_4$ flux. We can observe the presence of low intensity WZ spots (two of them are indicated by red arrows). Figure 3(a-d) shows TEM and HRTEM images (with [1-10] zone axis) of a typical self-catalyzed GaAs NW. With the applied growth conditions （T$_G$ = 600°C and V/III flux ratio = 2.3) the NW exhibits a pure-ZB phase almost over their entire length. Both ZB variants can be observed due to the presence of some twin planes (marked in green in Figure 3(a) and shown in Figure 3(d)). Then, near the NW top, we observe a sequence with: a transition zone about 60 nm in length with a high density of twin planes and some WZ segments (marked in blue in Figure 3(a) and shown in Fig. 3(c)), and finally a pure WZ segment about 100 nm in length followed by a thin ZB segment about 10 nm in length (marked in red in Figure 3(a) and shown in Figure 3(b)). This final sequence and the associated growth mechanism are well known thanks to previous studies[18–24,26]. Indeed, the GaAs NW growth is continued with the consumption the Ga droplet leading to a decrease of its volume. Recent in-situ TEM results obtained on Au- or Ga-catalyzed GaAs NWs[29–31] have confirmed that the nucleation of the ZB or WZ structure will occur depending on the droplet shape and more precisely on the wetting angle β which induces the location of the atomic layer nucleation. From these recent results, and as illustrated in Figure 4(a), the final structure sequence of the NWs can be explained as follow depending on β: the NW growth with a large wetting angle β, typically greater than 125°, will lead to the nucleation inside the droplet, giving thereby the growth of a ZB phase accordingly to the F. Glas model[7] (Figure 3(d) and Figure 4(a) stage i). Once the Ga shutter is closed, the droplet starts to be consumed leading to a decrease of its volume and so of β, hence leading to a nucleation at the TPL giving thereby the growth of a WZ phase, also accordingly to the F. Glas model[7]. This step leads first to a faulty section with twin planes and WZ/ZB phase mixture and then to a pure-WZ section (Figure 3(c and b) and Figure 4(a) stages ii to iv). Then, once the wetting angle becomes typically lower than ~55°, a final ZB segment is formed (Figure 3(b) and Figure 4(a) stage v).

The first purpose of this work was to follow this structure evolution via the RHEED pattern during the NW growth and at the growth end when the Ga shutter is closed. To proceed, RHEED videos were recorded along the $[1\bar{1}0]$ azimuth during the NW growth and the time-dependent intensities of the ZB and WZ spots were analyzed. The analyzed spots giving the ZB and WZ phase "intensities" are indicated in Figure 1. Figure 4(b) shows the evolution of the $\frac{I_{ZB}}{I_{ZB}+I_{WZ}}$ (or $\frac{I_{WZ}}{I_{ZB}+I_{WZ}}$) intensity ratio (IR), where $I_{ZB}$ and $I_{WZ}$ are the intensities of the ZB and WZ spots, respectively, as a function of the growth time. The green area corresponds to the NW growth with Ga and As flux (zone (i)). At the time t = 0 sec, WZ and ZB IRs are equal to 0.5, due to the absence of NWs (the measured intensities are those measured on the diffraction line of the

Si(111) substrate surface). When the Ga and As flux are opened, in a first time an increase of the ZB IR $\frac{I_{ZB}}{I_{ZB}+I_{WZ}}$ (respectively, a decrease of the WZ IR $\frac{I_{WZ}}{I_{ZB}+I_{WZ}}$), is observed (zone i1). However, after 90 seconds we can observe a short and slight decrease of the ZB IR (respectively, a short and slight increase of the WZ IR) for about 40 seconds (zone i2). After that, we see again an increase of the ZB IR (respectively, a decrease of the WZ IR), which tends to become constant with the growth time with a value close to 1 for the ZB IR (respectively close to 0 for the WZ IR - zone i3 and corresponding stage (i) in Figure 4(a)), meaning that the NWs are entirely or quasi-entirely ZB at this moment.

After 600 seconds, when the Ga shutter is closed, we can assume that the Ga droplet volume starts to decrease thus leading to the sequence schematized in the stages (i) to (v) of Figure 4(a). A delay is observable (zone (ii)) between the closing of the Ga shutter and the beginning of the WZ IR increase (respectively, the ZB IR decrease). This delay is interpreted as due to the consumption time of the droplet which is necessary to reach the wetting angle leading to the WZ phase (corresponding to stage ii in Figure 4(a)). This measured delay is about 20 seconds and it must match with the transition zone t with a high density of structural defects (observed in Figure 3(c)). Then, we observed an increase of the WZ IR (respectively, a decrease of the ZB IR) during about 70 seconds corresponding to the pure-WZ segment (zone (iii)) and to stages iii and iv in Figure 4(a)), and finally a weak decrease of the WZ IR (respectively, a weak increase of the ZB IR) corresponding to the growth of the final ZB segment (zone (iv) and to stage v in Figure 4(a)). A stabilization of both ZB and WZ IRs is finally observed after the end of the NW growth when the droplet is totally consumed (zone (v)). These RHEED measurements are those perfectly in line with the structural evolution of the NW observed by TEM measurements and that we can therefore associate with the size and wetting angle modifications of the Ga droplet as illustrated in Figure 4(a).

It should be noted that the observed RHEED diagram can be affected by the density or the length of the NWs due to shadowing effect or by the incident angle of the electron beam. However, the purpose of this work is to monitor the evolution of the RHEED diagram during the growth.

### 3.2. Zinc Blende / Wurtzite alternation

The capability to analyze the RHEED pattern and the crystal structure evolution of the GaAs NWs during the growth was used to control the formation of a WZ segment inside ZB NWs. The Ga flux was stopped to reduce the size and wetting angle of the Ga droplet during a short time in order to induce the growth of a WZ segment without a total consumption of the Ga droplet. For such a purpose and based on the previous results, after 240 seconds of ZB NW growth, the Ga shutter was closed for 60 seconds in order to avoid the total consumption of the droplet as analyzed from Figure 4(b). An illustration of the expected behavior of the Ga droplet is given in Figure 5(a), while the evolutions of ZB and WZ IRs as a function of the growth time are plotted in Figure 5(b). At the growth beginning the NWs exhibit a ZB structure (first green area). Then, after the Ga shutter closing at 240 seconds for 60 seconds (red area), the WZ segment growth is well observable with the increase of the $\frac{I_{WZ}}{I_{ZB}+I_{WZ}}$ intensity ratio. It can be noticed that after the Ga shutter opening the increase of the WZ IR did not stopped immediately. This delay corresponds to the Ga droplet refeeding which had not yet reached the wetting angle necessary for the transition to the ZB phase. In order to determine these durations more precisely, the first derivative of the WZ IR is plotted between 200 s and 400 s (Figure 6). The period without the Ga flux is represented with the red area. The ZB-WZ and WZ-ZB phase transitions are indicated by the vertical black lines, where the first derivative of the WZ IR is equal to zero, and are used to estimate the length of the ZB and WZ segments (see table 1). From Figure 6, we approximate the growth duration of the WZ segment (including the mixed

ZB – WZ segments) to $t_{WZ}$ = 69 s. In order to calculate the length of the ZB and WZ segments, an average axial growth rate $v$ = 1.7 nm/s was used (from the average length of the ZB+WZ segments measured on around twenty NWs and the growth time). A comparison between the calculated lengths and the average lengths obtained by TEM measurements on a typical NW (Figure 5(c)) is reported in the Table 1.

By extending the semi-empirical growth model for Au-catalyzed GaAs[25,32,42] and InAs[43] NWs, we proposed in Vettori *et al.*[44] a model for self-catalyzed GaAs NWs able to account for the droplet evolution as a function of (a) the direct Ga and As fluxes and (b) the Ga atoms entering the droplet by diffusion on the $SiO_2$-terminated Si substrate (for short NWs) and on the NW facets (more details can be found in the Supplementary Information section). Motivated by the droplet stability at a solid angle, an original feature of the model exposed by Vettori *et al*[44]. is the existence of an upper-limit wetting angle for the droplet. As a consequence, aside the classical axial NW growth the model is able to predict conditions that trigger both the axial and the lateral NW growth depending on the droplet evolution. In our case, we expect the numerical simulations to confirm the unique (critical) value of the wetting angle that characterizes the ZB-WZ and WZ-ZB phase transitions. Figure 7 shows the time-evolution for : (a) the Ga and As amounts of atoms feeding the droplet, (b) the droplet and NW radii, (c) the wetting angle and (d) the NW length. The closure of the Ga shutter between 240 sec and 300 sec induces a decrease of the droplet radius (visible in (b)) and of the wetting angle (visible in (c)). Moreover, the decrease of the wetting angle (or droplet volume) induces a decrease of the droplet capture surface for As and, as a consequence, a slightly lower axial grow velocity (visible in (d)). When the Ga shutter is opened at 300 sec the droplet radius and the wetting angle increase again. At 370 sec the maximum wetting angle is attained so that the NW radius increases in order to accommodate a higher droplet volume.

The vertical dotted lines in Figure 7 (c) and (d) correspond to t=260 sec, t=330 sec when the ZB-WZ and WZ-ZB transitions occur as indicated in Figure 6. The corresponding ZB and WZ segment lengths predicted by the numerical simulation are 450 nm and 106 nm, in good agreement with the TEM analysis of respectively 443 nm and 122 nm. The numerical method predicts values of the wetting angle equal to 126° for the ZB-WZ transition and 124° for the WZ-ZB transition, in agreement with results recently reported in Kim *et al.*[30] and measured by *ex situ* TEM. It should be noted that this critical wetting angle is close to the values reported for Au-catalyzed GaAs NWs[29,31].

## 4. Conclusion

During this work, the indexation of the ZB and WZ spots observed on the RHEED pattern and the determination of the time needed to entirely consume the droplet were realized as well as the establishment of a procedure capable to produce an extended WZ segment using only an *in situ* characterization system such as the RHEED system. The optimization of the growth temperature is still needed as well as the good management of the Ga flux to control the droplet shape. All the results were based on the *in situ* RHEED analysis and were validated and characterized using TEM imaging. Using numerical simulation for the NW growth and our *in situ* RHEED measurements, a critical wetting angle of about 125° for the ZB / WZ phase transitions was found.

**Acknowledgements:**

The authors thank the NanoLyon platform for access to equipments and J. B. Goure for technical assistance. The authors want also to acknowledge the French Agence Nationale de la

Recherche (ANR) for its funding to the ANR BEEP (project 18-CE05-0017-01) and to the ANR Equipex TEMPOS, which funded the Transmission Electron Microscope (project 10-EQPX-0050).

**Figures**

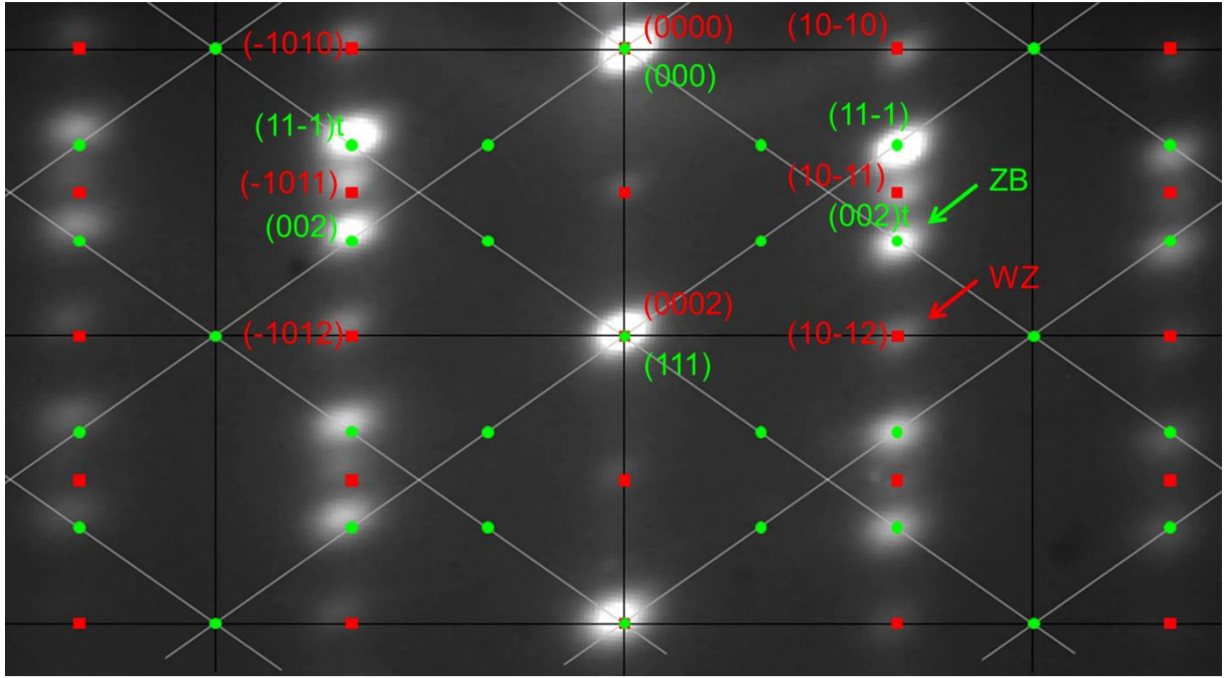

Figure 1: RHEED pattern recorded along the [1$\bar{1}$0] azimuth, where ZB and WZ spots are visible, superposed with an indexation diagram. Green dots report the ZB plans taking into account the two ZB variants. Red squares reports WZ plans. The ZB and WZ spots whose intensities are measured are indicated by the green and red arrows, respectively. An extinction of the structure factor $F_{hkl}$ is responsible of the missing spots of one over two ZB columns, and of the one over two WZ spots along the growth axis. The spots visible between the bright spots along the growth axis are present due to other orders of diffraction.

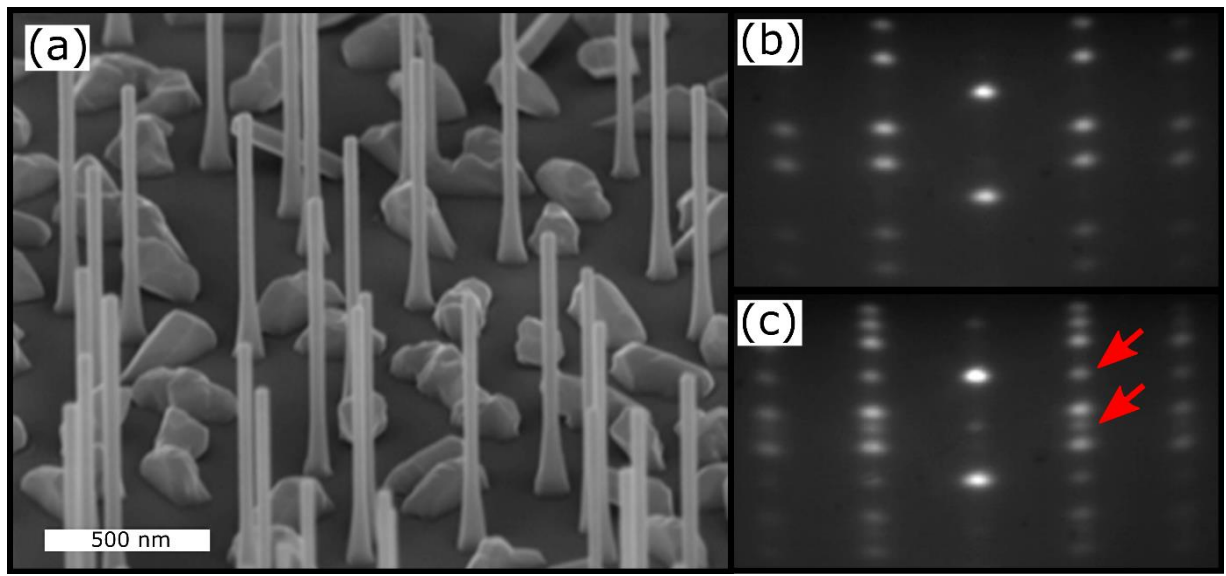

Figure 2: (a) SEM image of the measured sample. (b) RHEED pattern measured along the [1$\bar{1}$0] azimuth during the NW growth, where only ZB spots are visible. (c) RHEED pattern measured at the end of the NW growth after closing the Ga shutter, where WZ spots are visible (indicated by red arrows).

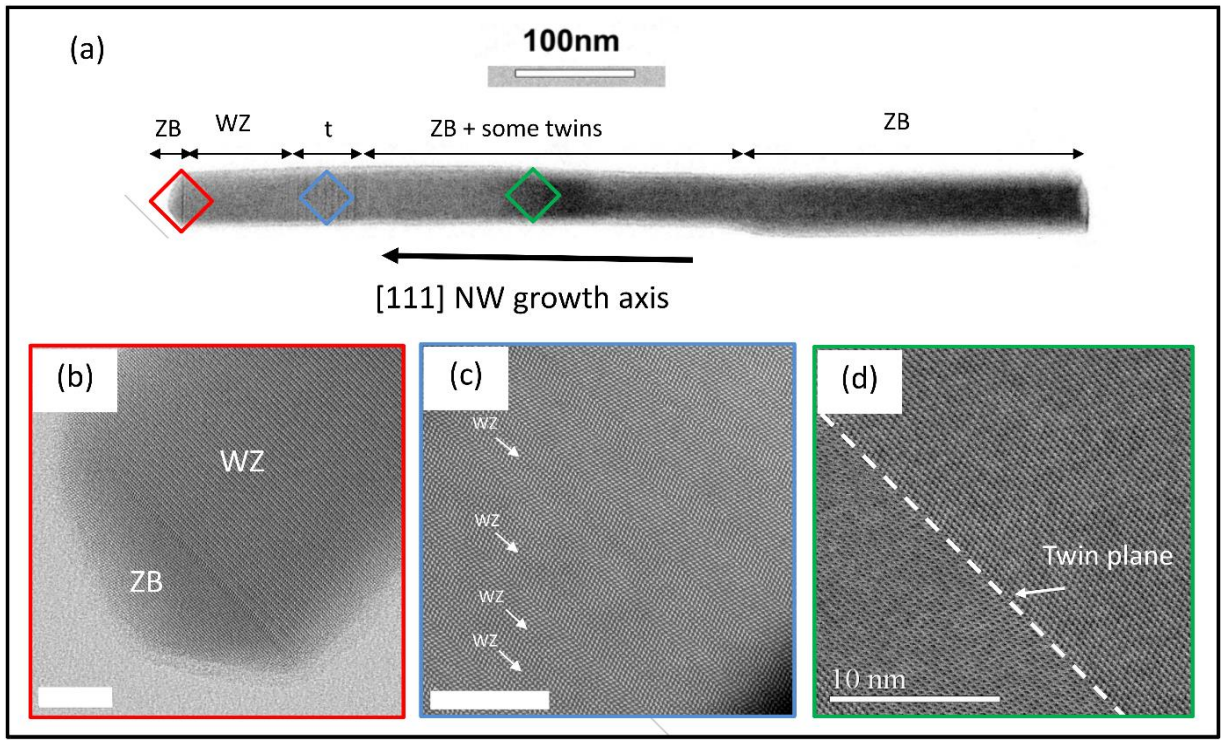

Figure 3: (a) (HR)-STEM image of a typical GaAs NW where different zones can be identified: (b) the small ZB segment grown at the end of the droplet consumption, (c) the transition zone t with a high density of twin planes and some WZ segments, (d) the ZB phase of the NW with some twin planes for length greater than ~ 200 nm. Scale bars in (b), (c) and (d) are 10 nm. (a), (b) and (d) are BF-STEM images while (c) is a HAADF-STEM image.

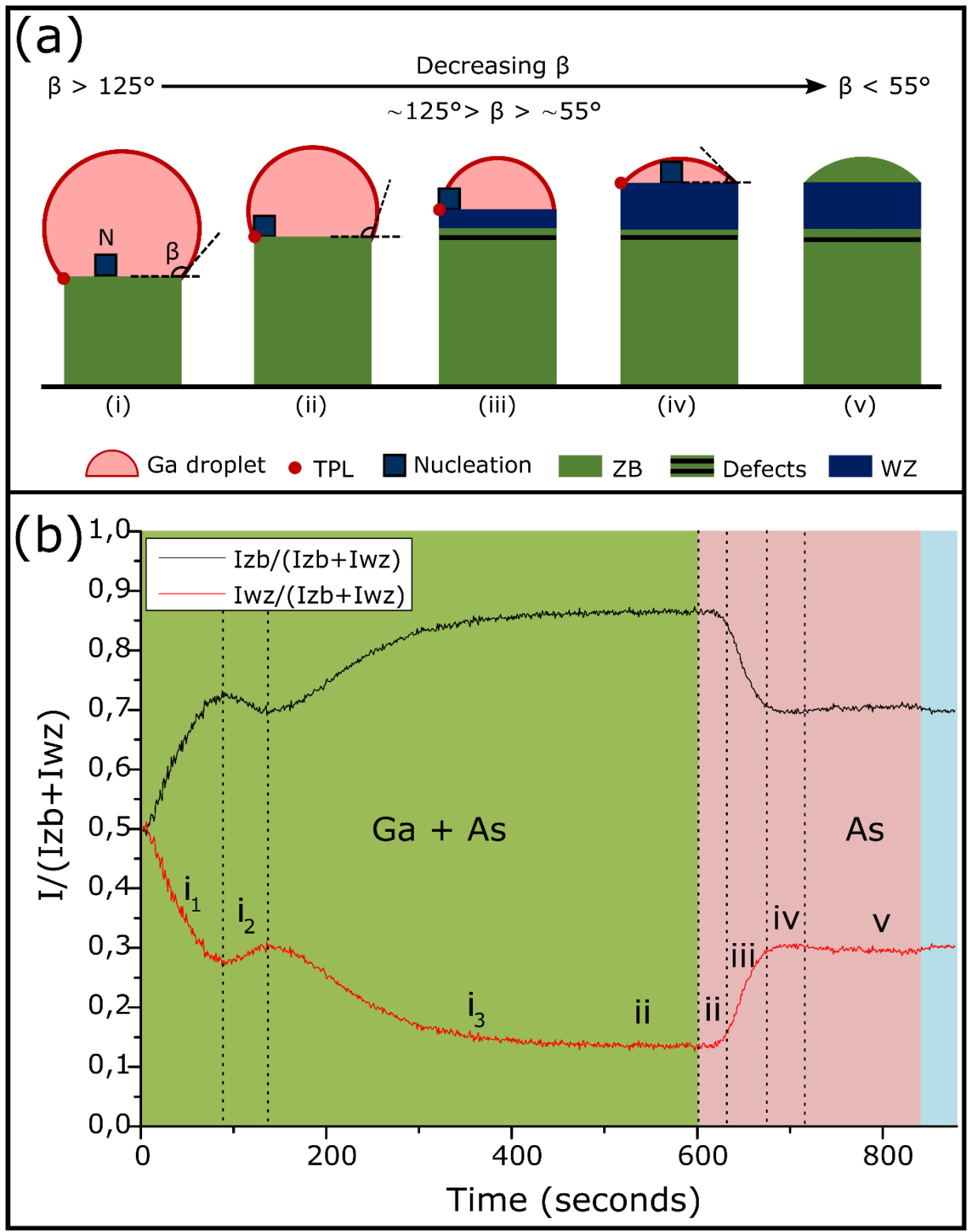

Figure 4 (a) Scheme of the Ga droplet evolution: (i) during the growth of the GaAs NWs with Ga and As flux leading to a ZB phase. Then, after closing of the Ga shutter, the droplet volume and wetting angle started to decrease leading to: (ii) a transition zone t with structural defects; (iii) the growth of the pure-WZ segment; (iv) the retreat of the droplet leading to a move of the nucleation on the NW top facet and to the growth of the final and short ZB segment (v). (b) $\frac{I_{ZB}}{I_{ZB}+I_{WZ}}$ and $\frac{I_{WZ}}{I_{ZB}+I_{WZ}}$ intensity ratios as a function of the growth time. In green the growth under Ga and As flux, in red the growth under As flux, and in blue the sample cooling under As flux.

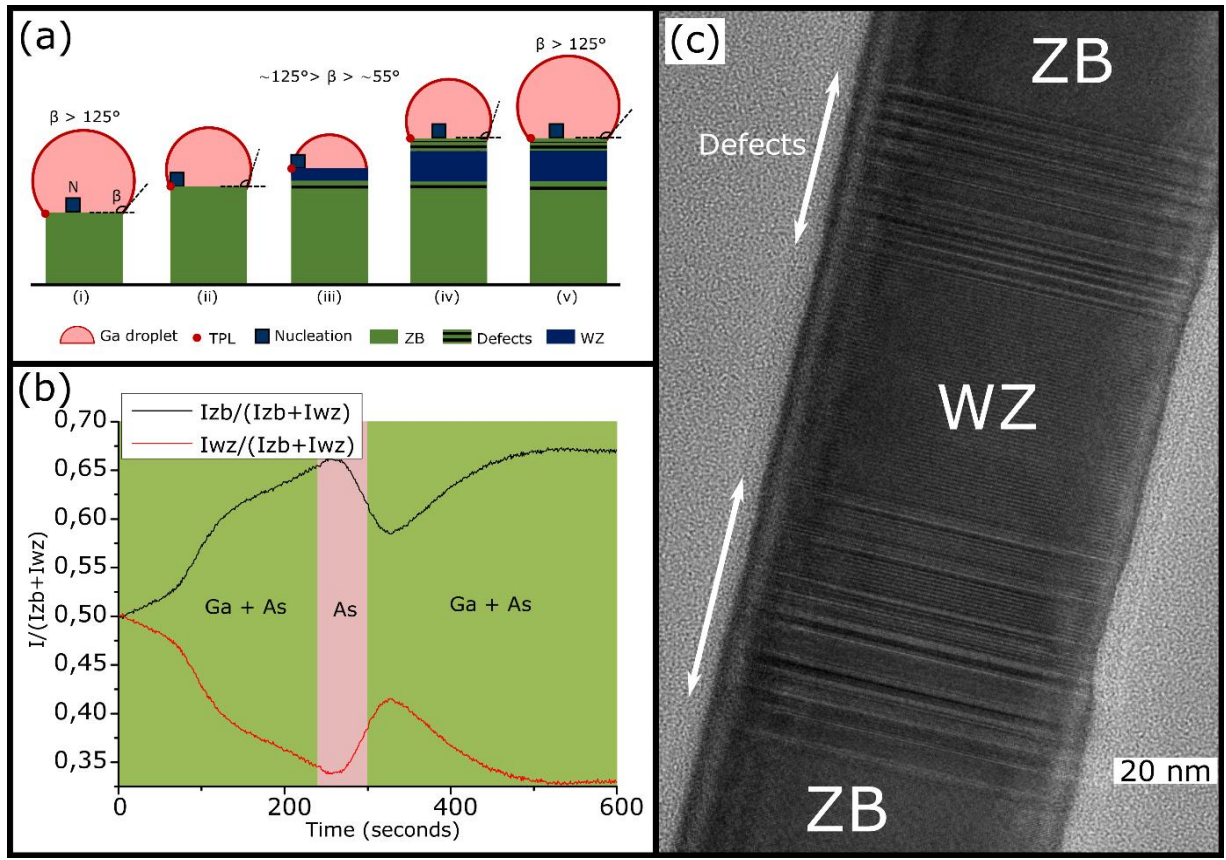

Figure 5: (a) Scheme of the expected Ga droplet volume evolution and corresponding crystal phases during the adopted procedure: (i) Growth of the ZB GaAs NW under Ga and As flux. (ii) Closing of the Ga shutter leading to a reduction of the droplet volume and thus of β leading to the formation of a defect section. (iii) Formation of a WZ section. (iv) Opening of the Ga shutter leading to the re-feeding of the droplet and thus to an increase of β leading to the formation of a defect section followed by (v) a ZB segment. (b) Intensity ratios $\frac{I_{ZB}}{I_{ZB}+I_{WZ}}$ and $\frac{I_{WZ}}{I_{ZB}+I_{WZ}}$ as a function of the growth time. In green the growth under Ga and As flux, in red the growth under As flux only. (c) HRTEM image showing the produced WZ segment in the ZB GaAs NW.

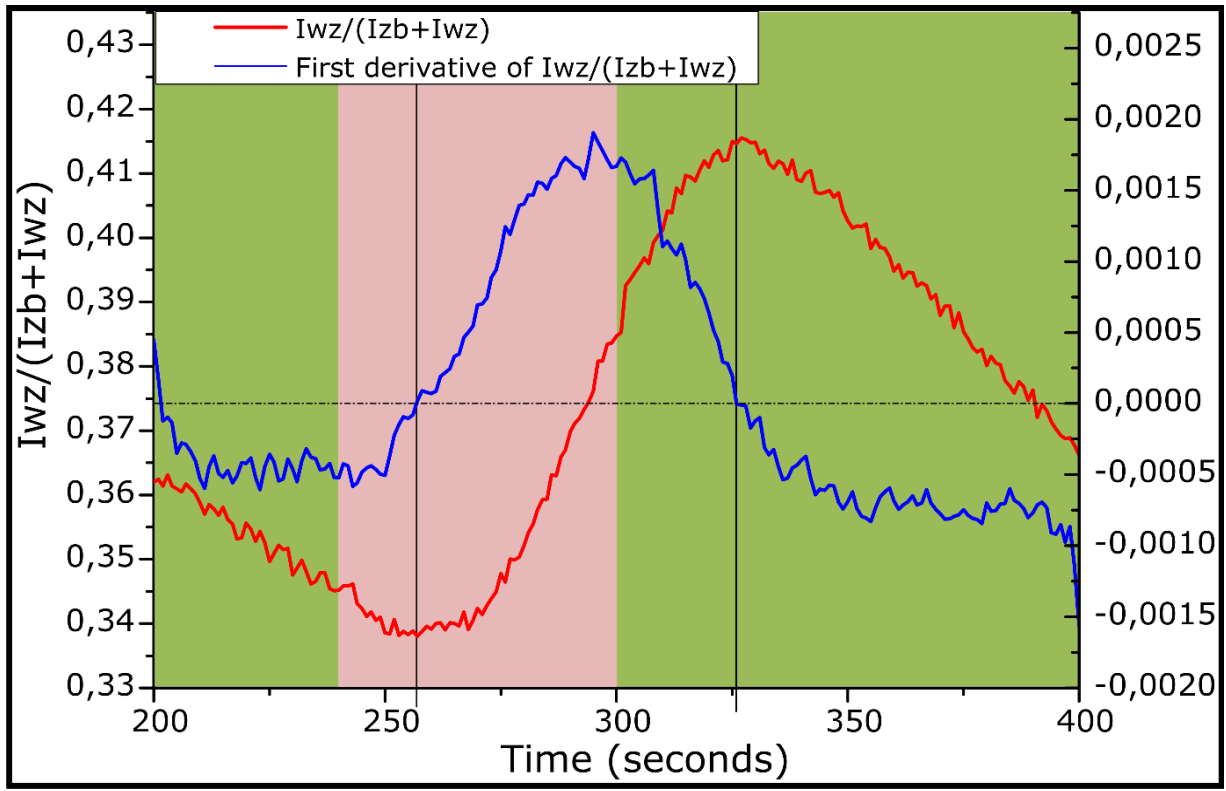

Figure 6: Zoom of the intensity ratio $\frac{I_{WZ}}{I_{ZB}+I_{WZ}}$ (red curve) corresponding to the WZ segment growth. Green areas correspond to the growth under Ga and As flux, red area corresponds to the growth under As flux only. The blue curve and the right axis are related to the first derivative of the WZ intensity ratio, while the dotted line corresponds to the zero of this curve. The transition times were extracted from the intersection between the dotted line and the first derivative.

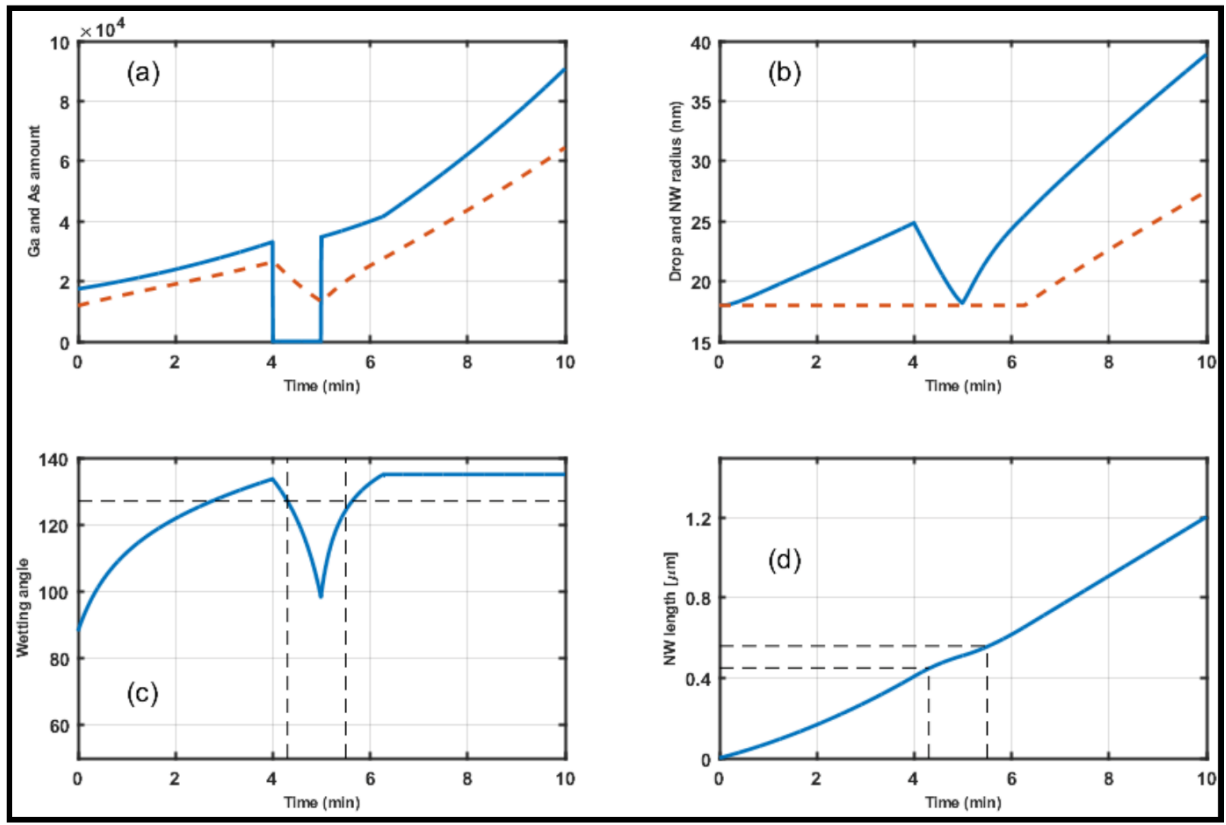

Figure 7: Time evolution for: (a) the amount of Ga (number of atoms in continuous blue line) and As (dashed red line) atoms feeding the droplet, (b) the droplet (continuous blue line) and NW (dashed red line) radii, (c) the wetting angle (in degrees) and (d) the NW length. The numerical results were obtained using the following numerical values: the Ga and As sources located at incidence angle equal to 27.9° and 41°, respectively, with respect to the normal to the substrate, nominal Ga and As fluxes given by: $F_{Ga}$= 3.53 atoms/sec.nm$^2$, $F_{As}$ = 27.3 atoms/sec.nm$^2$. The best fit was obtained using 38 nm and 1400 nm for the diffusion lengths on the SiO$_2$-terminated Si substrate and on the NW facets, respectively, and the As concentration threshold in the droplet is fixed at 1%. From experimental data, typical values for the initial conditions for the droplet are r =18 nm and wetting angle = 90°.

|  | Growth time (seconds) | Measured length (nm) (from TEM) | Calculated length (nm) (from RHEED) | Simulated length (nm) |
|---|---|---|---|---|
| Bottom ZB segment | 257 | 443 | 437 | 450 |
| WZ segment | 69 | 122 | 117 | 106 |

Table 1: Reported values of the growth times, the average measured lengths extracted from TEM analysis, the calculated lengths obtained using RHEED measurements (with an average growth rate of 1.7 nm/s) and the simulated lengths. The growth time was obtained using the Figure 6. The measured lengths are average values obtained by measurements on TEM image of around twenty NWs. The frontier between the ZB and the WZ segment are defined from the first derivative of the WZ IR for RHEED analysis. The WZ segment measured by TEM includes the stacking faulted areas.